\documentclass[twocolumn, twocolappendix]{aastex631}
\usepackage{amsmath}

\newcommand{\vx}{\mathbf x}
\newcommand{\vk}{\mathbf k}

\newcommand{\rev}[1]{#1}

\begin{document}

\title{Recovering Cosmic Structure with a Simple Physical Constraint}

\author[0009-0001-6592-6467]{Tian-Cheng Luan}
\affiliation{School of Physics and Astronomy, Sun Yat-Sen University, No.2 Daxue Road, Zhuhai, 519082, China}

\correspondingauthor{Xin Wang, Xiao-Dong Li}
\email{wangxin35@mail.sysu.edu.cn, lixiaod25@mail.sysu.edu.cn}


\author[0000-0002-2472-6485]{Xin Wang}
\affiliation{School of Physics and Astronomy, Sun Yat-Sen University, No.2 Daxue Road, Zhuhai, 519082, China}
\affiliation{CSST Science Center for the Guangdong-Hong Kong-Macau Greater Bay Area, SYSU, China}

\author{Jiacheng Ding}
\affiliation{Shanghai Astronomical Observatory, Chinese Academy of Sciences, No. 80 Nandan Road, Shanghai, 200030, China}

\author{Qian Li}
\affiliation{School of Physics and Astronomy, Sun Yat-Sen University, No.2 Daxue Road, Zhuhai, 519082, China}
\affiliation{CSST Science Center for the Guangdong-Hong Kong-Macau Greater Bay Area, SYSU, China}

\author{Xiao-Dong Li}
\affiliation{School of Physics and Astronomy, Sun Yat-Sen University, No.2 Daxue Road, Zhuhai, 519082, China}
\affiliation{CSST Science Center for the Guangdong-Hong Kong-Macau Greater Bay Area, SYSU, China}

\author{Weishan Zhu}
\affiliation{School of Physics and Astronomy, Sun Yat-Sen University, No.2 Daxue Road, Zhuhai, 519082, China}
\affiliation{CSST Science Center for the Guangdong-Hong Kong-Macau Greater Bay Area, SYSU, China}



\begin{abstract}

Radio observation of the large-scale structure (LSS) of our Universe faces major challenges from foreground contamination, which is many orders of magnitude stronger than the cosmic signal. While other foreground removal techniques struggle with complex systematics, methods like foreground avoidance emerge as effective alternatives. However, this approach inevitably results in the loss of Fourier modes and a reduction in cosmological constraints. We present a novel method that, by enforcing the non-negativity of the observed field in real space, allows us to recover some of the lost information, particularly phase angles. We demonstrate that the effectiveness of this straightforward yet powerful technique arises from the mode mixing from the non-linear evolution of LSS. Since the non-negativity is ensured by mass conservation — one of the key principles of the cosmic dynamics — we can restore the lost modes without explicitly expressing the exact form of the mode mixing.  Unlike previous methods, our approach utilizes information from highly non-linear scales, and has the potential to revolutionize the analysis of radio observational data in cosmology. Crucially, we demonstrate that in long-baseline interferometric observations, such as those from the Square Kilometre Array (SKA), it is still possible to recover the baryonic acoustic oscillation (BAO) signature despite not directly covering the relevant scales. This opens up potential future survey designs for cosmological detection. 

\end{abstract}



\section{Introduction} \label{sec:intro}

Radio observation of large-scale structures (LSS) holds significant potential for extracting cosmological information about our Universe. In particular, the redshifted 21cm emission line from neutral hydrogen is a powerful tool for mapping late-time LSS and is a primary focus of many ongoing and future radio telescopes, including CHIME \citep{chime2022}, Tianlai\citep{Tianlai2015ApJ}, HIRAX \citep{HIRAX2016SPIE}, \rev{CHORD \citep{2019clrp.2020...28V}} and SKA \citep{SKA2020PASA} etc. Unlike galaxy redshift surveys that are capable of resolving individual galaxies, LSS surveys in the radio band typically employ a technique known as intensity mapping (IM). Over the past decade, this method has been validated multiple times via cross-correlating the radio dataset \citep{Chang2010Natur,Masui2013} with the galaxy surveys \citep{Drinkwater2009}. 
\rev{Additionally, stacking analyses of CHIME data have detected cross-correlations between IM and samples of galaxies and quasars \citep{2023ApJ...947...16A}.}
\rev{More recently, ref.~\cite{Paul2023} claimed a measurement of the auto-power spectrum at small scales using the MeerKAT array. If confirmed, this would highlight its potential for future cosmological applications.}

Despite significant advancements in recent years, 21cm intensity mapping still faces considerable challenges in achieving its ultimate goal of providing robust cosmological constraints. The main obstacle is the contamination from the foreground, which is about four orders of magnitude greater than the cosmological signals. 
Many strategies have been developed to mitigate its impact, including methods leveraging spectral smoothness \citep{Santos2005ApJ,Liua2009MNRAS,Petrovic2011}, dominance of the foreground (Principal Component Analysis, PCA) \citep{Masui2013,switzer2013,switzer2015,2015MNRAS.454.3240B}, \rev{the non-Gaussianity (Independent Component Analysis, ICA) \citep{Chapman2012,wolz2014,2017MNRAS.464.4938W}, the Karhunen-Loève transform \citep{RShaw2014ApJ,RShaw2015PhRvD,2022PhRvD.105h3503B}, and methods based on machine learning \citep{2020MNRAS.494..600M,2021ApJ...907...44V,2021JCAP...04..081M,2021ApJ...916...42W} etc.}
Particularly, the spectral smoothness indicates that the foreground primarily occupies the lower wavenumber region along the line of sight in Fourier space, where $k_{\parallel} < k_{\mathrm int}$, with $k_{\mathrm int}$ denoting the wavenumber threshold of this `intrinsic' foreground. 
Moreover, in real observation, the chromatic instrumental response can further lead to \rev{frequency-dependent fluctuations on beam \citep{2024ApJS..274...44D}} and a mixing of foreground modes with the HI signal within a wedge-like region of the Fourier space
$ k_{\parallel} < \left [ D_c(z) H(z) \theta_0/(c (1+z)) \right] k_{\perp} $ \rev{\citep{2020PASP..132f2001L}},  where $\theta_0$ could either be defined as the angular size of primary beam or more conservatively the size of the horizon, $D_c(z)$ is the comoving distance to redshfit $z$, $H(z)$ is the Hubble parameter at $z$.
Instead of directly tackling the foreground contamination, a more conservative approach is to avoid these Fourier regions where the foreground resides, known as foreground avoidance (FA). While this strategy effectively eliminates foreground interference, it also leads to a loss of information, potentially impacting cosmological measurements such as baryonic acoustic oscillations etc.

\rev{
For a Gaussian random field, different modes are independent, making information lost in a specific Fourier region irretrievable. } However, the highly nonlinear nature of structure formation leads to a coupling between different modes. 
This well-known fact has led to the development of various algorithms that can recover lost information in LSS. 
For example, the tidal reconstruction method \citep{zhu2016} constructs a quadratic estimator from the tidal field to extract large-scale modes from small-scale perturbations. Similar techniques have been successfully applied to extract late-time lensing signals from CMB observations \citep{Maniyar2021} as well. 
\rev{In addition, with the rapid advancement of machine learning techniques, significant efforts have been made to leverage various neural networks to extract higher-order coupling information and recover wedge modes \citep{2021MNRAS.504.4716G, 2024MNRAS.529.3684K, 2024arXiv240721097S,LiQian2024arXiv}.}
However, the quadratic estimator relies on the second-order perturbation theory. The regime where we can apply such a technique is limited. Meanwhile, mode coupling is very efficient at a highly nonlinear regime, where all modes carry traces of those missing larger-scale modes. While extending the quadratic estimator to higher-order perturbations is feasible, the resulting formalism would be exceedingly complex. 
\rev{On the other hand, while machine learning approaches show promise and can effectively utilize information from highly nonlinear regimes, they are typically computationally demanding and require substantial effort to incorporate survey-specific instrumental systematics into their training process.  
Consequently, developing a realistic algorithm that operates in the deeply nonlinear regime and is directly applicable to real observations remains technically challenging.
}

\rev{In this paper, we introduce a remarkably simple yet effective procedure to restore lost information, particularly phase angles. This approach is based on the fact that the large-scale structure of the universe evolves under mass conservation, ensuring a strictly positive density. The removal of foreground-contaminated modes disrupts this positivity, but enforcing it helps recover the missing information. Thus, this method offers a promising alternative for addressing this challenge and could have a significant impact on future survey designs.
This paper is organized as follows. Section \ref{sec:toy_model} presents the core concept using a simple one-dimensional toy model. Section \ref{sec:cosmo_apply} presents the main results based on three-dimensional cosmological simulation data. Finally, Section \ref{sec:discussion} provides further discussion. }

\section{Phase Recovery with a Toy Model of Structure Formation} \label{sec:toy_model}

To begin, we would like to remind the reader a seemingly obvious fact, that the cosmic density field of dark matter (DM) is inherently non-negative.  Although this condition may appear trivial, it has significant implications for the nonlinear evolution of  structure formation. At early times, the nearly Gaussian primordial density perturbation, $\delta_0 \ll 1$, causes the density field $\rho/\bar{\rho} \approx 1+\delta_0 $ to be `almost always' \footnote{Neglecting the long tail of the Gaussian distribution. Of course, strictly speaking, the physical requirement of non-negativity implies that there exists a slight non-Gaussianity in  $\delta_0$.} non-negative.  
As the density perturbation gradually grows, non-negativity is maintained through mass conservation, one of the two fundamental equations governing the nonlinear evolution of density perturbations. 
In Fourier space, this non-linearity is expressed through recursive mode coupling as $\delta^{\mathrm nl}(\vk) \propto \delta_{\mathrm D}(\vk-\vk_1-\vk_2) \alpha(\vk_1, \vk_2)\delta(\vk_1) \delta(\vk_2)$ \citep{Bernardeau2002PTreview}, where $\alpha$ is the mode-coupling coefficient and $\delta_{\mathrm D}$ is the Dirac delta function.
This means that missing modes from the foreground contamination are coupled with the remaining Fourier modes. The new insight we have here is that such coupling is closely related to the non-negativity condition in the real space and enforcing such requirement could potentially help us restore the information lost in the foreground wedge.

Specifically, consider the observed temperature map $T_\mathrm{kcut}$, in which all Fourier modes within the foreground wedge have been removed. 
Compared to the true non-negative temperature map\footnote{Unlike the 21cm signal from the epoch of reionization, the late-time HI budget is primarily concentrated in galaxies, and the non-negativity condition still applies to the temperature map $T(\vx)$. }, $T_{\mathrm kcut}$ exhibits negative values in some locations due to the absence of wedge modes. This indicates that {\it the locations of these negative regions actually contain valuable information about those missing modes}. 
To harness this information, we propose an extremely simple process in this paper, which we term `positivization'.\footnote{In signal processing, a similar procedure is known as rectification. However, due to the distinct purpose and application here, we refer by a different name.} Essentially, this process enforces the non-negativity condition by highlighting those negative regions and setting them to zero.  This is equivalent to applying a window function $F$ in configuration space, so that the `positivized' field 
\begin{eqnarray}
    T_\mathrm{posi} (\vx) = F\left[T_\mathrm{kcut}\right] (\vx) ~T_\mathrm{kcut}(\vx).
\end{eqnarray}
\rev{Here the window $F\left[T_\mathrm{kcut} \right] (\vx)$ is the Heaviside step function, which depends on the values of the foreground-removed temperature map $T_\mathrm{kcut}$, i.e. $F\left [T_\mathrm{kcut}\right] =\left \{ 1 \left(T_\mathrm{kcut} \ge 0\right); ~ 0 \left(T_\mathrm{kcut}<0 \right) \right \}$. Although this expression may make the process appear linear, it is, in fact, a non-linear operation, as $F$ itself depends on the value of $T_\mathrm{kcut}$.}

\begin{figure*}
    \centering
    \includegraphics[width=1.0\textwidth]{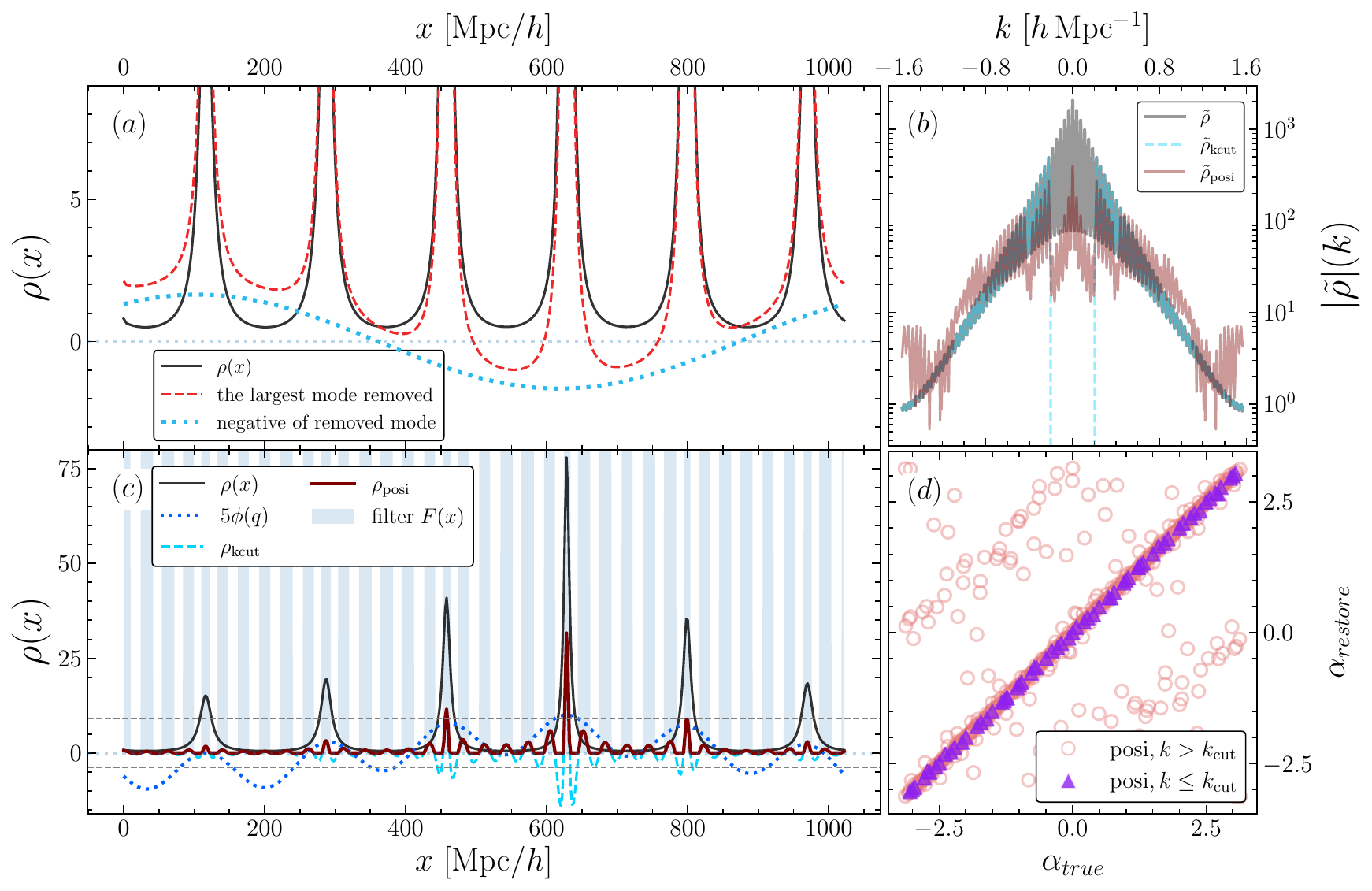}
    \caption{
    Demonstration of the `positivization' process in a one-dimensional toy universe with two non-zero modes of primordial displacement fluctuation $\phi(q)$.
    Panel (a) illustrates the core concept of positivization following the removal of the largest mode (not counting the zero mode) from the density field. The red dashed line represents the density after mode removal, while the black solid line depicts the original density field, and the light blue dotted line shows the negative of the removed mode. Notably, the locations of the negative values in the mode-removed field align with the peak of the missing mode.    Panel (b) illustrates the Fourier amplitude of various fields, demonstrating that after positivization, some power re-emerges in the removed region, albeit with an amplitude approximately one order of magnitude lower (brown lines).
    Panel (c) displays a similar process conducted after subtracting all Fourier modes with $k<0.2~ h\,\mathrm{Mpc^{-1}}$. 
    The non-linearly evolved density is depicted as the black solid line, while the density field with modes removed, $\rho_\mathrm{kcut}$, is shown as a sky blue dashed line. The positivized field, depicted as a brown solid line, results from setting negative values of $\rho_\mathrm{kcut}$ to zero, with the corresponding filter $F[\rho_\mathrm{kcut}](x)$ displayed as light blue bars. 
    Most crucially, panel (d) shows a scatter plot of the phase angle $\alpha$ between the true and restored fields. As indicated by the purple triangles, the phase of the missing modes aligns almost perfectly with the true phases. However, the positivization process introduces a systematic shift in the phase of the remaining modes, resulting in a bias, as shown by the red circles. 
    }
    \label{FIG:LSS1d}
\end{figure*}

\begin{figure*}
    \centering
    \includegraphics[width=1.0\textwidth]{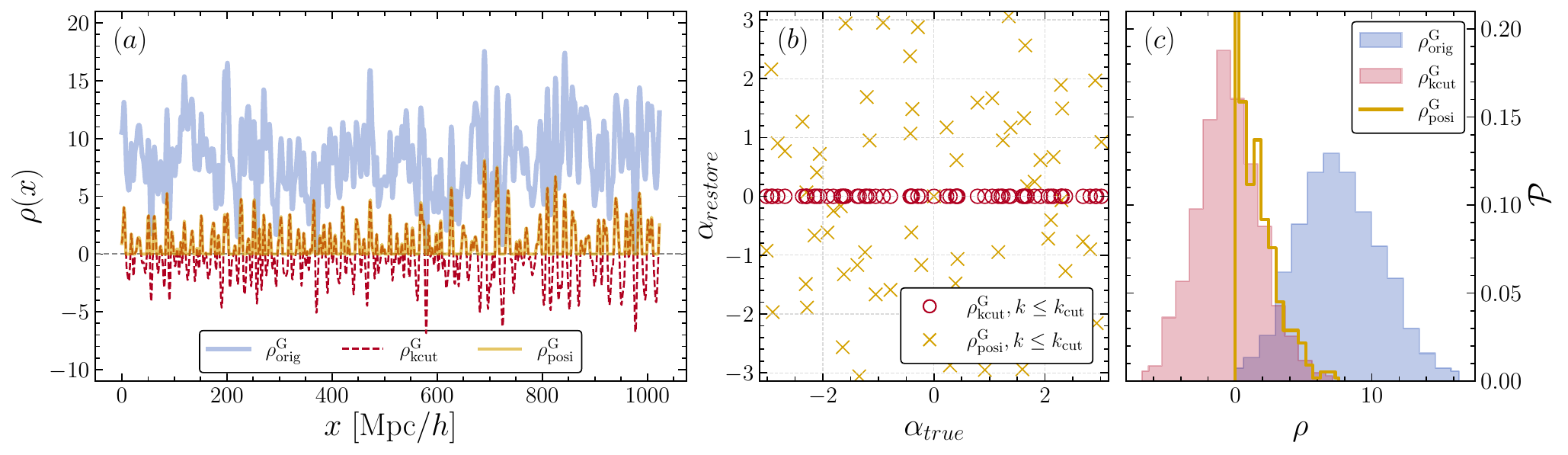}
    \caption{
    \rev{Demonstration of the `positivization' procedure applied to a one-dimensional positive Gaussian random field.
    In panel (a), The thick blue solid line represents the original Gaussian field, $\rho^\mathrm{G}_\mathrm{orig}$, the red dashed line shows the field after removing modes with $k< 0.2~ h\,\mathrm{Mpc^{-1}}$, $\rho^\mathrm{G}_\mathrm{kcut}$, and the golden solid line depicts the positivized result, $\rho^\mathrm{G}_\mathrm{posi}$.
    Panel (b) displays the phase angle $\alpha$ between the true and reconstructed fields in the large-scale region ($k\leq k_\mathrm{cut}$). The red circles correspond to $\rho^\mathrm{G}_\mathrm{kcut}$, where the large modes are set to zero, while the golden crosses correspond to $\rho^\mathrm{G}_\mathrm{posi}$, the positivized result. The phase angles in the positivized field are randomly distributed and show no correlation with the original phases, demonstrating that the positivization procedure relies on non-linear mode coupling to recover the phase angles.
    Panel (c) shows the probability density function (PDF, $\mathcal{P}$), where the blue histogram represents the original field, the red histogram corresponds to $\rho^\mathrm{G}_\mathrm{kcut}$, and the golden histogram line illustrates the value redistribution after positivization.}
    }
    \label{FIG:LSS1d_gauss}
\end{figure*}

To better grasp the physical intuition behind the positivization process, let us first explore a heuristic toy model of structure formation. In this simplified universe, we consider only two non-zero modes of fluctuations \footnote{A single mode universe also works but is less effective in mode-coupling.}
\begin{eqnarray}
    \phi(q) =  \sum_i^{N} \tilde{\phi}(k_i) \exp\left[ \mathrm{i} \left( k_i q+\alpha_i\right)\right], 
\end{eqnarray}
where $\phi(q)$ and $\tilde{\phi}(k)$ represent the primordial displacement potential in real and Fourier space, respectively. Here, $k$ denotes wave vector of the Fourier mode, and $q$ refers to the Lagrangian coordinate, $\alpha$ is the complex phase of each mode, with the total number of modes $N=2$. 
The displacement potential $\phi(q)$ under consideration is depicted as the blue dotted line in panel (a) and (b) of Fig.~\ref{FIG:LSS1d}. To enhance visibility of the fluctuation, $\phi(q)$ is amplified by a factor of five. Applying the Zel'dovich approximation, the density distribution at a later epoch can be described by the mass conservation equation
\begin{eqnarray}
\label{eqn:rhoza}
    \rho(x) = \left( \frac{\partial x}{\partial q} \right)^{-1}
    =\left[ 1+ a(t) \frac{d^2\phi}{dq^2} \right]^{-1}
\end{eqnarray}
The result is shown as the solid black line in panel (a) sof Fig.~\ref{FIG:LSS1d}. The box size is set as $1024 ~ \mathrm{Mpc/h}$, consistent with cosmological context usually considered. In this example, we did not specify a particular cosmic time $t$; instead, we simply chose the value for $a$ that offers a visually reasonable representation of the non-linear density $\rho(x)$. 
Due to non-linear mode coupling, the initial two-mode fluctuation evolves into a density distribution with a complex Fourier power spectrum, as depicted by the solid grey lines in panel (b). Consequently, all these Fourier modes of $\rho(x)$ are somewhat interconnected and correlate with each other. Therefore, the non-negativity condition, a direct consequence of mass conservation, is closely related to this interconnectedness of the Fourier modes. 

\begin{figure*}
     \centering
    \includegraphics[width=1.0\textwidth]{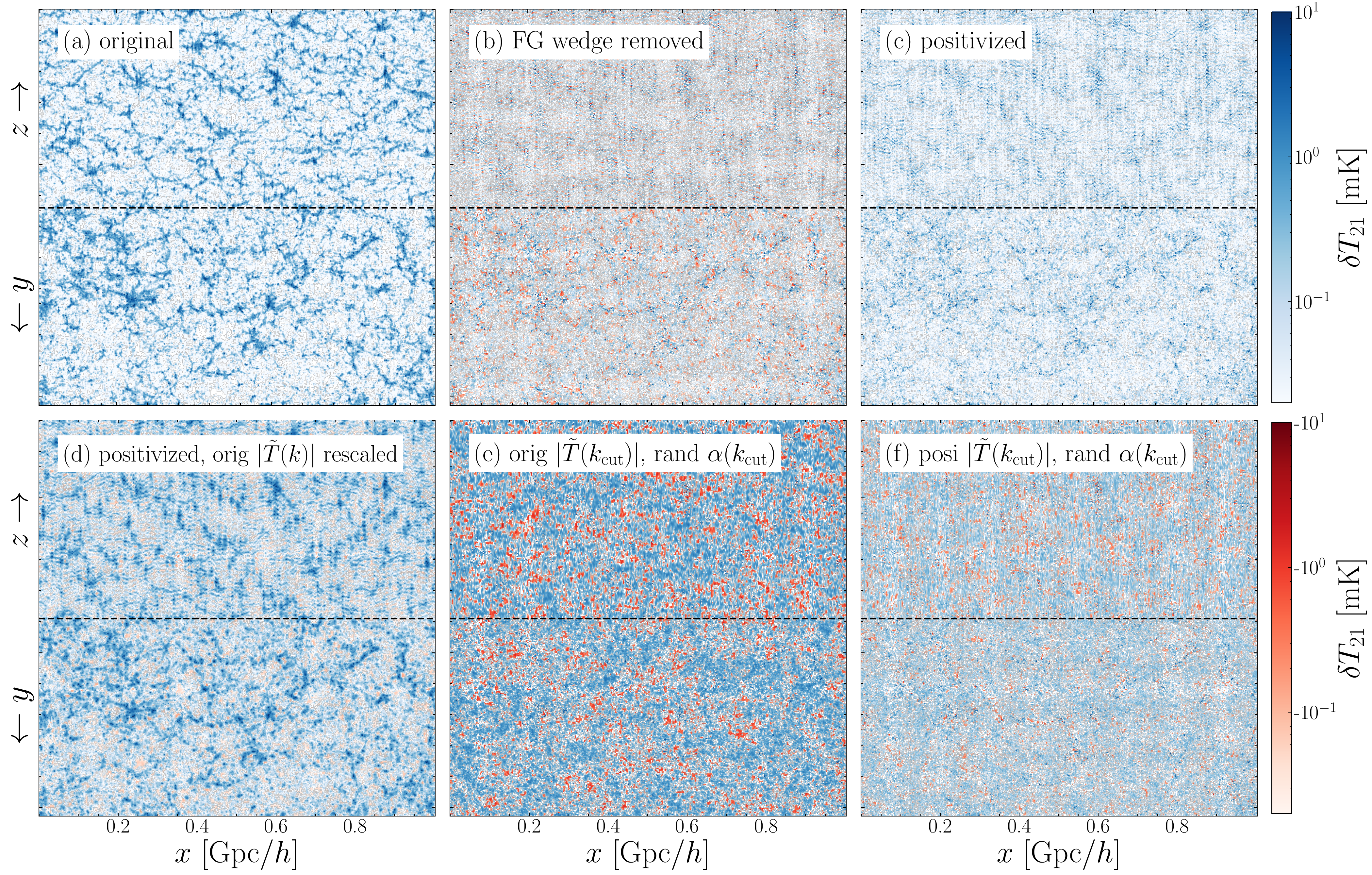}
    \caption{
    Slice of a simulated 21cm intensity map in redshift space at $z=1$, with the $z$ axis representing the line of sight direction. The slice has a thickness of approximately $1.95~\mathrm{Mpc}/h$. To better illustrate the visual changes in redshift space, each panel is divided into two sections: the upper part displays the $x-z$ plane, while the lower part shows the $x-y$ plane. The color bar uses blue to indicate positive values and red for negative values. Panel (a) displays the original 21cm temperature map, $T_{\mathrm orig}$, while panel (b) illustrates the `observed' map, $T_{\mathrm kcut}$, where Fourier modes within the foreground wedge are lost. To demonstrate the effectiveness of our approach, we selected a very conservative Fourier region as being contaminated, $k_{\parallel} < \mathrm{max}(0.5 ~h\,\mathrm{Mpc^{-1}}, \;0.3k_{\perp})$ (also see Fig.~\ref{FIG:Ck2d} for the shape of this region).  We then apply the `positivization' process, resulting the field in panel (c). Additionally, with prior knowledge of the power spectrum, the amplitude of Fourier modes in the positivized map, $T_{\mathrm posi}$, can be further rescaled. The results, shown in panel (d), remarkably resemble the original map displayed in panel (a). 
    To demonstrate phase recovery of the positivization process, panel (e) displays a map where the phase angles of the contaminated modes have been randomized, yet the Fourier amplitude is matched to the true $T_{\mathrm orig}$. Similarly, panel (f) illustrates a phase-randomized map with the Fourier amplitude matching with $T_{\mathrm posi}$. 
    }
    \label{fig:simslice}
\end{figure*}

To simulate the foreground contamination and avoidance strategy in a simplified  setting, we remove only the largest mode of density $\rho(x)$ (not counting the zero mode). The outcome is illustrated by the red dashed line in panel (a). The negative value of the removed fluctuation is shown as blue dotted line. As a result, this subtraction leads certain areas of the original $\rho(x)$, depicted as the black solid line, to display negative values. 
It is crucial to note that {\it the location of the negative region aligns with the missing mode}, indicating that this location retains the phase information of the missing mode.\footnote{\rev{Of course, an underlying assumption is that the removed modes have sufficiently large amplitudes to drive some regions into negative values. In our specific application in radio cosmology, this assumption does hold, as the power spectrum in the foreground-contaminated region is generally higher.}}
Therefore, employing a strategy that utilizes these negative regions can, to some extent, recover the phases of these missing modes.

To investigate more realistic scenarios, we then remove all Fourier modes with $k<0.2~ h\,\mathrm{Mpc^{-1}}$. The result, denoted as $\rho_{\mathrm kcut}$, is displayed as the sky blue dashed line in panel (c). In this case, the amplitude of high-density `halos' is substantially reduced. As before, it leads to regions that once again display negative values.  
Following the positivization procedure, we set all negative values to zero, as illustrated by the brown solid line in the same panel. The light blue bars indicate the locations of positive regions, corresponding to the filter $F(x)$. After this process, some power was restored, as depicted by the brown lines in panel (b), though the amplitude is about an order of magnitude lower.
To examine the phase restoration, we present the scatter plot comparing the phase angle $\alpha$  between $\rho_{\mathrm posi}$ and the original $\rho$, as shown in panel (d). The phases of restored modes ($k\le k_{\rm cut}$), indicated by purple triangles, align almost perfectly with original phases. However, this process also impacts the phases of the remaining modes (shown as red circles), causing a systematic phase shift.

\begin{figure*}
    \centering
    \includegraphics[width=1.0\textwidth]{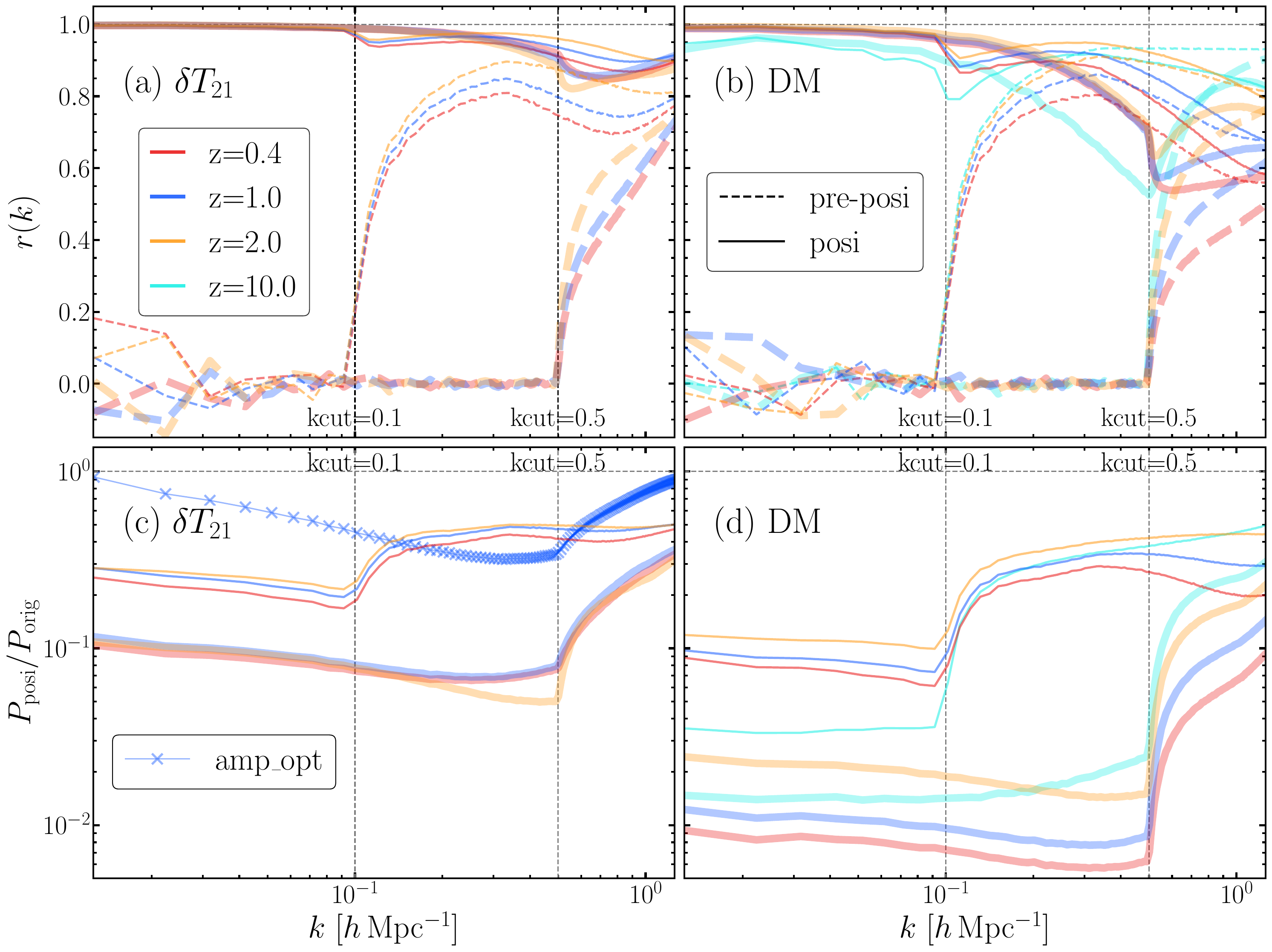}
    \caption{One dimensional cross-correlation ratio $r(k)$ (panel a and b) and power spectrum ratio (panel c and d) between various processed fields and the original temperature or dark matter density field.
    Different colors represent results for various redshifts, ranging from $z=0.4$ to $z=2$ and $z=10$. 
    Dashed lines in $r(k)$ panels indicate the cross-correlation with the pre-positivized field, $\rho_{\mathrm kcut}$. Thicker solid lines show results after applying positivization to fields affected by a larger foreground wedge ($k_{\parallel} < \mathrm{max}(0.5, \; 0.3k_{\perp})$), where more Fourier modes have been removed. Thinner solid lines correspond to fields with a smaller affected region ($k_{\parallel} < \mathrm{max}(0.1, \; 0.3k_{\perp})$). Although the phases of these restored Fourier modes remain highly correlated with the true values, their amplitudes differ significantly. Panel (c) and (d) present the transfer function $T(k)^2 = P_{\mathrm{posi}} / P_{\mathrm{orig}}$, defined as the ratio of the power spectra between the positivized and original fields for $\delta T_{21}$ and the dark matter density field, respectively. For $k_{\mathrm cut}=0.5~ h\,\mathrm{Mpc^{-1}}$, the power spectrum of the positivized temperature field is approximately one order of magnitude lower than the original field, with variations depending on the cutoff scales, while for the dark matter field, the ratio is about two orders of magnitude lower. Additionally, the blue cross line shows the amplitude-optimized result (see Appendix \ref{apdx:amp_opt}). 
    }
    \label{FIG:Ck1d}
\end{figure*}

\begin{figure*}
    \centering
    \includegraphics[width=1.0\textwidth]{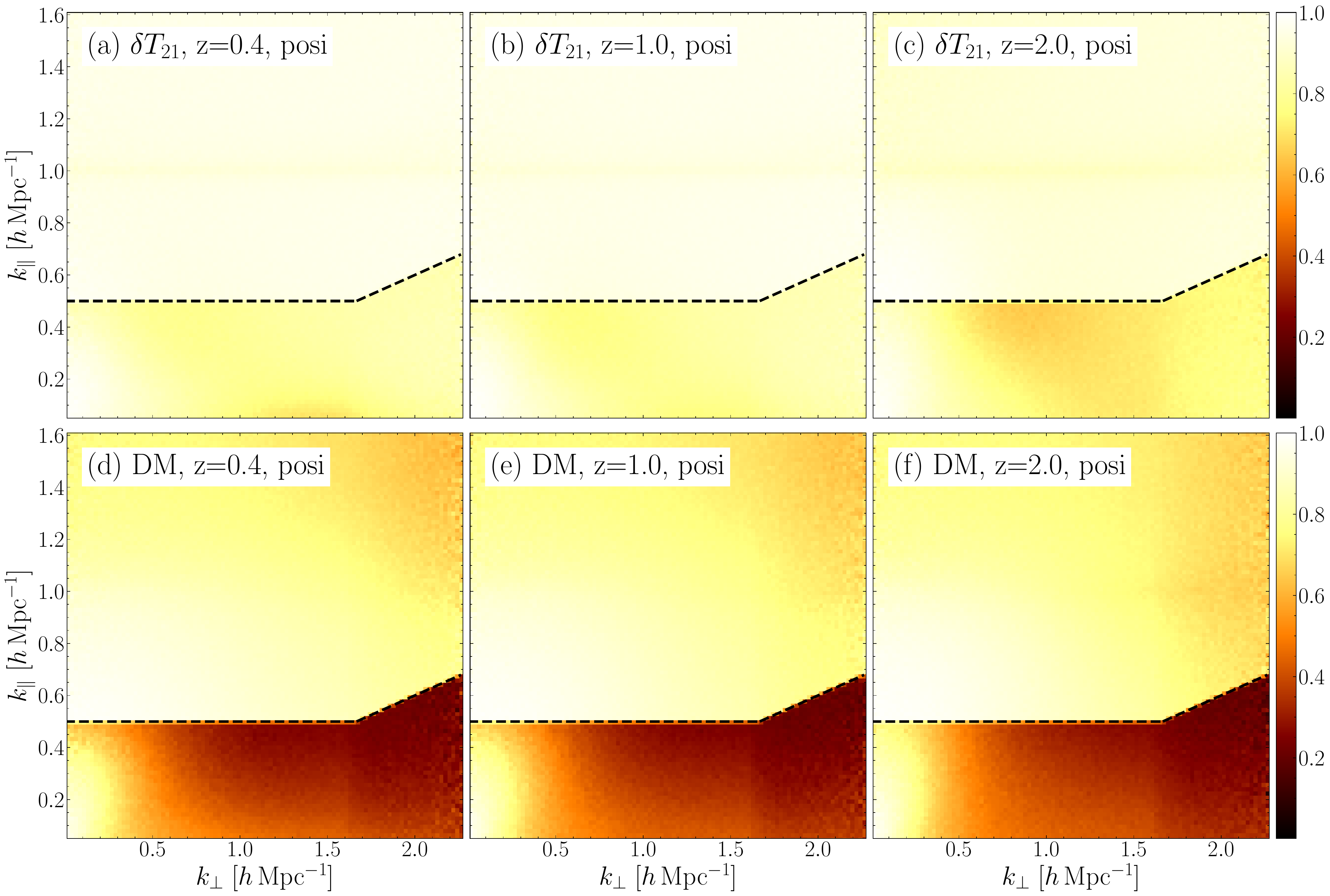}
    \caption{Illustration of the two-dimensional cross-correlation ratio \(r(k_{\perp}, k_{\parallel})\) between the positivized $\delta T_{21}$ (upper panels) and dark matter (lower panels) with the original field across different redshifts. In each panel, the black dashed line marks the boundary of foreground wedge. Here, we only demonstrate the more aggressive removal of foreground $k_{\parallel} < \mathrm{max}(0.5, \;0.3k_{\perp})$. 
    }
    \label{FIG:Ck2d}
\end{figure*}

\rev{To further investigate the role of nonlinear mode coupling, we applied the same procedure to a positively defined Gaussian random field, shown as the blue solid line in panel (a) of Figure \ref{FIG:LSS1d_gauss}. After removing Fourier modes with $k<0.2~ \mathrm{h~ Mpc^{-1}}$, we obtained the red dashed line, and after applying positivization, the result is shown as the yellow solid line. 
As seen in panel (b), the newly emerged phase angles (yellow crosses) are randomly distributed and show no correlation with the original phases. We also illustrate the one-dimensional probability density functions of the three fields in panel (c). This result is expected, as the Fourier modes of a Gaussian random field are statistically independent, meaning that any information lost in a specific Fourier region is irretrievable.
Thus, the positivization procedure is only effective for non-linearly coupled non-Gaussian fields, as it relies on the nonlinear mode coupling inherent in LSS evolution.
}

\section{Cosmological Simulation and Application} \label{sec:cosmo_apply}

Given the results observed in our toy universe, how might this procedure perform 
under more realistic conditions? To explore this question, we use a simulated 21cm temperature map generated with N-body simulation, onto which an analytic HI-mass relation \citep{Villaescusa-Navarro2018} is applied to model the HI distribution and subsequently the brightness temperature map. For further details on the simulation methodology, please refer to Appendix \ref{apdx:sim}. 
Panel (a) of Fig.~\ref{fig:simslice} presents a slice of the true redshift space temperature map in $x-z$ (upper) and $x-y$ (lower) plane, with $z$ axis representing the line of sight direction. Panel (b) then illustrates the observed map $T_{\mathrm kcut}(\vx)$ after removing all Fourier modes within the foreground wedge. To demonstrate the effectiveness of our method, we have selected an extremely conservative region of the wedge, with $k_{\parallel} < \mathrm{max}(0.5 ~ h\,\mathrm{Mpc^{-1}}, \;0.3 k_{\perp})$. 
Compared to the true distribution in panel (a), the absence of certain Fourier modes significantly reduces the amplitude of the observed temperature in panel (b), leading to many regions becoming negative, which are highlighted in red.

Following the demonstration in the toy model, we applied the positivization process by setting all negative values to zero. The resulting field is displayed in panel (c) of Fig.~\ref{fig:simslice}. As shown, despite the complications introduced by redshift space distortion along the line-of-sight direction in the upper panel, a faint resemblance to the original field in the $x-y$ slice can still be observed. Keep in mind that the structural similarity occurs when the phase information is preserved. Moreover, with prior knowledge of the power spectrum, the amplitude of Fourier modes in the positivized map, $T_{\mathrm posi}$, can be further rescaled. The results, shown in panel (d), remarkably resemble the original map displayed in panel (a). For comparison, panels (e) and (f) display the same field with the power spectrum identical to the original (panel a) and the positivized fields (panel c), respectively, but with randomized phase information in the foreground wedge. These comparisons clearly show that the cosmic web morphology is partially retained in the positivized field.

Furthermore, we present in the first two panels of Fig.~\ref{FIG:Ck1d} the one-dimensional cross-correlation ratio between the original distribution and the positivized field, defined as $r(k) = P^{\mathrm posi, orig}(k)/\sqrt{P^{\mathrm posi} (k) P^{\mathrm orig} (k) }$. Here, $P^{\mathrm posi, orig}$ is the cross power spectrum between the positivized and original fields, and $P^{\mathrm posi}$ and $P^{\mathrm orig}$ are their respective auto power spectra. The results are shown as solid color lines in the first panel for the temperature map and the second for the dark matter distribution. 
Here we display results for two different wedge regions defined by $k_{\parallel} < \mathrm{max}(k_{\mathrm cut}, \;0.3k_{\perp})$, featuring a realistic threshold with $k_{\mathrm cut} =  0.1~h\,\mathrm{Mpc^{-1}}$ and a more aggressive threshold of $k_{\mathrm cut}=0.5~h\,\mathrm{Mpc^{-1}}$.  
In both cases, the cross-correlation ratio of the positivized field within the wedge region approaches unity but decreases to approximately $0.9$ around the threshold scale, marked by vertical lines. 
For comparison, the correlation ratio for the observed $T_{\mathrm kcut}$, shown as dashed lines, is near zero in the same region, consistent with the removal of information in this area.
Lines of different colors represent results at various redshifts. For the dark matter field, the recovered $r(k)$ is higher at lower redshifts, consistent with the expectation that fields at lower redshifts are more nonlinear and exhibit tighter coupling. However, this difference is less pronounced in the temperature map.  Furthermore, the phase recovery in the 21cm temperature map generally outperforms that of the dark matter field. This may be due to the higher non-linearity of the temperature field. 
In panel (c) and (d) of Fig.~\ref{FIG:Ck1d}, we demonstrate that although the phase information may be recovered, the amplitude of these Fourier modes is notably reduced. Here, we display the transfer function $T^2(k)=P_{\mathrm posi}/P_{\mathrm ori}(k)$ as the ratio of power spectrum between the positivized and the original field. As illustrated, the power spectrum of the positivized temperature map is approximately one order of magnitude lower than that of the original field, and for the dark matter field, it is about two orders of magnitude lower. Furthermore, the degree of amplitude reduction is also influenced by different redshifts and cut-off scales. 
This suggests that, by using a prior theoretical model of the power spectrum, one could rescale each Fourier mode to more accurately reconstruct the original map. Panel (e) of Fig.~\ref{fig:simslice} shows a visual slice of this rescaled field, revealing a significantly enhanced visual similarity to the original field. 

\begin{figure*}[!htbp]
    \centering
    \includegraphics[width=1.0\textwidth]{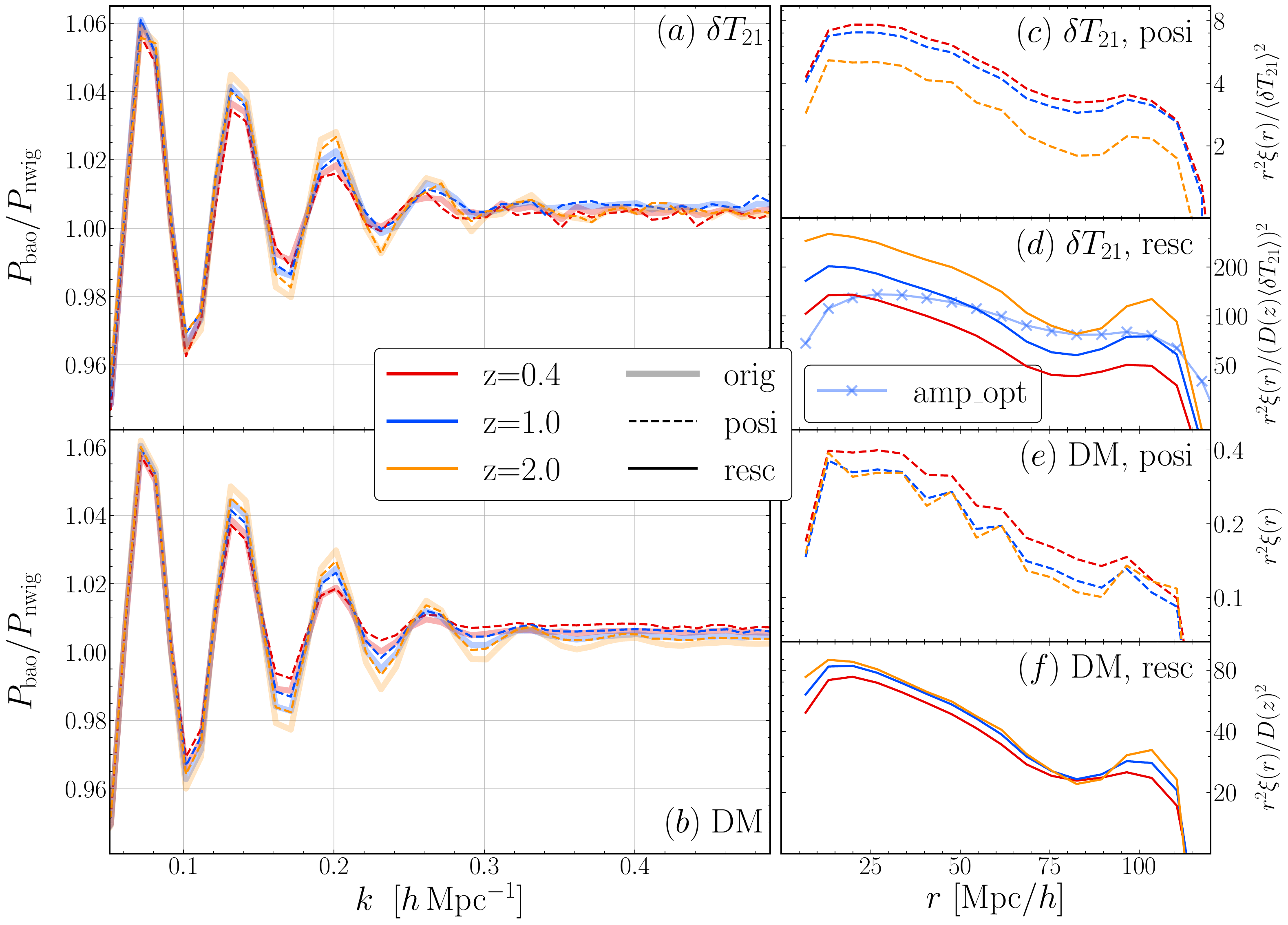}
    \caption{
    Demonstration that the positivization process preserves BAO signals in both the $\delta T_{21}$ (panel a, c and d) and Dark matter fields (panel b, e and f). 
    In all panels, red, blue, and orange lines represent results at redshifts $z=0.4$, $1.0$, and $2.0$, respectively. Thick lines denote the original simulation (orig), dashed lines indicate the signal after positivization (posi), and thinner solid lines show the positivized field with Fourier amplitudes rescaled to match the original power spectrum (resc).
    The left two panels (a and b) display the ratio of power spectra between simulations with and without BAO in their initial conditions, $P_{\mathrm{bao}} / P_{\mathrm{nwig}}$. For both the temperature and dark matter fields, the wiggles of the positivized field, whether with or without amplitude rescaling, closely resemble the original signal. 
    Furthermore, panels (c) to (f) display the two-point correlation function of various fields. Notably, panels (c) and (e) demonstrate that the BAO bump is also discernible in the positivized temperature and dark matter fields. And as shown in panel (d) and (f), the BAO bump is much clearer if one further rescales the amplitude of the missing modes. Notice that, due to different time evolution and amplitude reduction, each panel has been scaled accordingly. In addition, in panel (d), the blue cross line shows the amplitude-optimized result (see Appendix \ref{apdx:amp_opt}).  
    }
    \label{FIG:BAO}
\end{figure*}

In Fig.~\ref{FIG:Ck2d}, we also illustrate the cross-correlation ratio in two-dimensional $k_{\parallel} - k_{\perp}$ space. Similar to the one-dimensional results, the 21cm temperature map exhibits a higher correlation ratio. In this case, the phase information is almost perfectly recovered for the selected redshifts $z=0.4$ and $2$. For the dark matter field, the cross-correlation ratio is notably higher at lower $k$ values and gradually decreases at high $k$ within the foreground wedge.

Given these results, one might wonder about the potential applications of this simple process. Specifically, can the Baryon Acoustic Oscillation (BAO) signal be detected in the positivized field? Concerns may arise due to the reduced Fourier amplitudes of the positivized field. However, as illustrated in the left panels of Fig.\ref{FIG:BAO}, the BAO signature is indeed preserved. This is demonstrated by comparing the power spectra ratios of positivized simulations with and without BAO in their initial conditions, labeled as `BAO' and `nwig' respectively. We can observe the BAO in this way because the transfer function of the process, as shown in panel (c) and (d) of Fig.~\ref{FIG:Ck1d}, is a smooth function. 
It is important to emphasize that, as a conservative measure, we have removed all BAO signals by applying a cutoff scale of $k_{\mathrm cut} = 0.5~ {h\,\mathrm Mpc^{-1}}$. This means that even if all Fourier modes within the BAO region are lost due to foreground contamination, the signature can still be reconstructed through nonlinear coupling effects. Of course, these two panels (a and b) does not reflect the typical approach to measuring the BAO signature in practice. To provide a more realistic analysis, we plot the two-point correlation function in the right panels (c-f) of Fig.~\ref{FIG:BAO} without the help of those `no-wiggle' simulations. Due to different time evolution and amplitude reduction, we have re-scaled the correlation function accordingly. 
In panels (c) and (e), we demonstrate that the BAO bump can indeed be detected in the correlation function of the positivized field, despite a significant reduction in the amplitude of large-scale Fourier modes. Notably, the BAO signal in the temperature map is more prominent than in the dark matter map, due primarily to a less significant reduction in amplitude.

Furthermore, as shown in panel (d) and (f), assuming prior knowledge of large-scale Fourier amplitudes enhances the clarity of the BAO signal. However, in practice, accurately predicting the amplitude, particularly the non-linear power spectrum of the brightness temperature map, can be challenging. 
Given that the recovered amplitude is known to be reduced, it is possible to further numerically optimize the amplitude using the non-negativity condition. We detail this method in the Materials and Methods section and illustrate the results as the blue crossed line in panel (d) of Figure \ref{FIG:BAO}.

\section{Discussion} \label{sec:discussion}

In this paper, we introduce a novel technique for recovering the missing Fourier modes that are lost to foreground contamination. Despite its simplicity, the non-negativity condition is a fundamental constraint underlying the complex nonlinear structure formation process, offering an efficient method for restoring these lost modes. Unlike previous techniques such as tidal reconstruction, our approach utilizes information deep in the highly non-linear regime, providing greater flexibility in its application. This proves particularly beneficial as the foreground wedge obscures a significant portion of small scales where perturbative methods are ineffective.

The timing of this method's proposal is timely, given the persistent challenges faced in 21cm intensity mapping experiments. These challenges are notably severe in measuring the auto-power spectrum, where existing foreground removal techniques struggle with various sky-based and instrumental systematics, including cross-talk suppression and frequency-dependent beam variations. Recent advancement \cite{Paul23Meerkat} indicates that employing foreground avoidance strategy in interferometric observations (e.g., MeerKAT and the upcoming SKA) provides a viable approach for detecting the 21cm intensity mapping auto-power spectrum, although at a much smaller scales than those typically associated with BAO. However, as demonstrated in this paper, it is still possible to recover the BAO signature by extracting information via non-linear mode coupling. This is extremely valuable as it opens up the potential future survey designs for cosmological detection. The caveat, however, is that simply enforcing the non-negativity condition recovers only the phase information, not the amplitude. Therefore, caution must be exercised before it is applied directly to future cosmological measurements.

This BAO restoration is still limited by various statistical and systematical errors. Since the restored large-scale modes are those that interact with local small-scale fluctuations, the cosmic variance cannot be smaller than what the survey volume allows. Here, the volume is determined by the frequency bandwidth along the line of sight (LOS) direction, and the baseline resolution in the perpendicular direction, $\Delta k_{\perp}$. While the frequency bandwidth is primarily set by the instrument's design, the perpendicular scale is decided by the field of view (FoV) at a single pointing, which depends on the shortest baseline and the size of individual antennas. At lower redshifts, this can limit the perpendicular scale below the BAO scale. Fortunately, techniques like mosaicing allow us to extend this size effectively. Note that modes outside the survey volume are not fully `observed' in their entirety, yet they still interact with local modes, uniformly affecting everything within the surveyed area. Consequently, this interaction introduces a systematic error when applying the non-negativity condition.

Additionally, the shot noise and instrumental noise within the observed region further affect the recovery of these modes. \rev{As illustrated in Fig.~\ref{FIG:noise}, introducing thermal noise significantly affects the performance of the positivization procedure, influencing the cross-correlation ratio $r(k)$ (panel a), power spectrum (panel b and d), and two-point correlation function (panel c and e). For example, in panel (a), when the instrumental noise level at the same level of the 21cm signal $\sigma_\mathrm{noise}=\sigma_\mathrm{21}$, the cross-correlation ratio at the cut-off scale ($k_{\rm cut}=0.5~h\,\mathrm{Mpc^{-1}}$) decreases from approximately $0.9$ to $0.7$, while reducing the noise level by half yields much better results.  
Moreover, panels (b) and (c) show that increasing the noise level also suppresses the amplitude of the recovered modes, requiring a larger rescaling factor to match a theoretical power spectrum. As a result, the noise is amplified by an even greater factor, further complicating the effective application of this technique without significantly longer integration times.
Similarly, residual small-scale foregrounds could have a significant impact. However, as their characteristics remain poorly understood, the extent of their influence is unclear. Given the importance of these issues, a more comprehensive study of noise effects and potential mitigation strategies is needed, which we aim to explore in future work.
}

Another relevant application is the reconstruction of BAO. The non-linear evolution of large-scale structures smears the BAO peak in the correlation function, diminishing its constraining power on the sound horizon scale and, ultimately, on dark energy. A standard approach involves reversing this process to `linearize' the BAO signature. However, the effectiveness of this approach can be significantly compromised by missing modes because accurate reconstruction requires precisely estimating particle displacements in real space, which are heavily influenced by these lost modes.  Therefore, our approach could also enhance further data processing to improve cosmological constraints.

\begin{acknowledgments}
X.W. expresses deep gratitude for the productive discussions with Ue-Li Pen. This work is supported by the National SKA Program of China (Grants Nos. 2022SKA0110200, 2022SKA0110202 and 2020SKA0110401), the National Science Foundation of China (Grants Nos. 12473006, 12373005), the China Manned Space Project with No. CMS-CSST-2021 (B01, A02, A03).
\end{acknowledgments}

%

\vspace{5mm}


\software{\texttt{GADGET-4} \citep{gadget2021},  
          \texttt{ROCKSTAR} \citep{rockstar2013}, 
          \texttt{CAMB} \citep{2011ascl.soft02026L},
          \texttt{PyTorch} \citep{NEURIPS2019_bdbca288},
          }



\appendix

\section{Simulation}\label{apdx:sim}

\begin{figure*}
    \centering
    \includegraphics[width=1.0\linewidth]{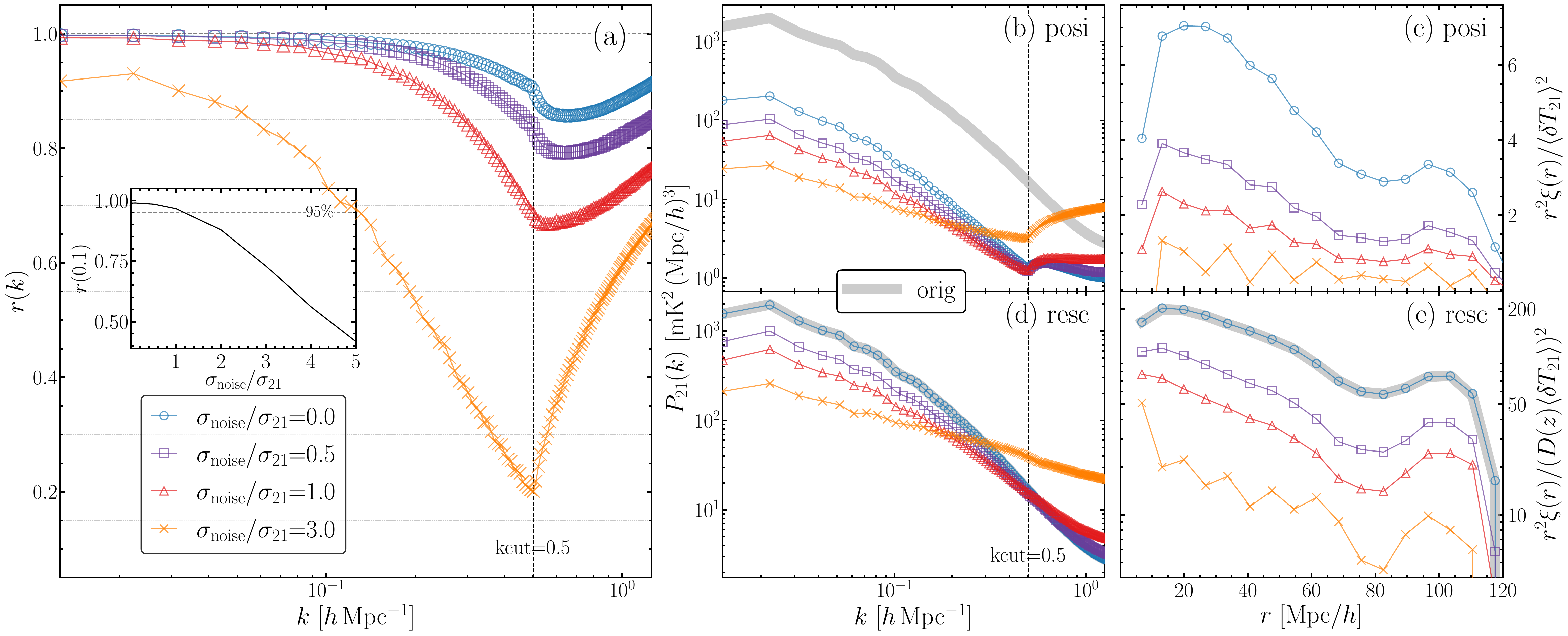}
    \caption{\rev{The impact of different noise levels $\sigma_{\mathrm{noise}}$ on the performance of the positivization process. For simplicity, we generate Gaussian random noise in configuration space as a pixel-by-pixel independent realization, with its variance defined relative to the 21cm signal variance $\sigma_{\rm 21}$. Specifically, we consider noise levels of $0$ (noiseless, blue circles), $0.5 \sigma_{\rm 21}$ (purple squares), $1\sigma_{\rm 21}$ (red triangles), and $3\sigma_{\rm 21}$ (orange crosses), respectively. Panel (a): The impact on the cross-correlation ratio $r(k)$ between positivized and the original fields. As expected, an increasing noise level gradually reduces the correlation ratio for the restored modes. For a more detailed depiction of noise dependence, the inset in panel (a) illustrates the correlation coefficient $r(k)$ at $k=0.1~h\,\mathrm{Mpc^{-1}}$ as a function of the noise level $\sigma_{\mathrm{noise}} / \sigma_{21}$. 
    Panels (b) and (d) illustrate the impact of noise on the restored power spectrum: panel (b) shows the results without rescaling, while panel (d) presents the results after rescaling with the same factor derived from the noiseless field. As shown, higher noise levels reduce the amplitude of the recovered modes, indicating that rescaling the amplitude to match a prior theoretical power spectrum would further amplify the noise. 
    Panel (c) and (e) demonstrate the similar positivized and rescaled results for the two-point correlation function. As noise levels increase, the amplitude gradually decreases, while the correlation function becomes noisier, progressively washing out the BAO peak. }
    }
    \label{FIG:noise}
\end{figure*}

Throughout this paper, we adopted the $\Lambda$CDM model with the cosmological parameters that match those reported by Planck 2018 \citep{planck2020}. N-body simulations were carried out using the publicly available open-source code \texttt{GADGET-4} \citep{gadget2021}. Our cosmological simulations adopted a box size of $1000~ \mathrm{Mpc}/h$,  containing $512^3$ particles. The initial conditions were generated using transfer function calculated by \texttt{CAMB} \citep{2011ascl.soft02026L} and second-order Lagrangian perturbation theory (2LPT) at redshift $z = 49$.
For 21cm intensity mapping observation, we first construct the map of neutral hydrogen from the N-body dark matter distribution using the prescription described by ref.~\cite{Villaescusa-Navarro2018} and then convert to the brightness temperature of 21cm signal. 
At lower redshifts, neutral hydrogen primarily resides within galaxies; however, the minimum halo mass threshold $M_{\rm hard}$ is below the halo mass resolution of our N-body simulation. Consequently, our HI map consists of two separate components: resolved halos and a subgrid contribution.

\textbf{Resolved halos:} The halo catalog  is obtained using the \texttt{ROCKSTAR} halo finder, which employs a 6D phase-space friends-of-friends algorithm \citep{rockstar2013}. The real-space positions of halos are converted to redshift-space by incorporating the LOS velocity of each halos
    \begin{equation}
        s = x_{\mathrm{los}} + \frac{v_{\mathrm{los}}}{aH(a)}. 
        \label{eq:RSD}
    \end{equation}
    Here, $x_{\mathrm{los}}$ and $v_{\mathrm{los}}$ represent the position and LOS velocity of halos, $a$ is the scale factor, and $H(a)$ is the Hubble expansion rate at $a$. The HI mass within a halo at redshift $z$, $M_{\rm{HI,Halo}}(M_{\rm{halo}}, z)$, is determined using fitting formula from \cite{Villaescusa-Navarro2018}, with parameters interpolated to the target redshift. We then assign $M_{\rm{HI,Halo}}$ to the nearest grid cell based on the halo's position in redshift space.

\textbf{Subgrid:} For halos below the mass resolution of our simulation, $M_{\rm halo, res}$, we begin with the redshift-space dark matter density distribution, obtained using the Cloud-in-Cell (CIC) method from the positions and line-of-sight velocities of dark matter particles. 
    In a grid cell with volume $V_{\rm{cell}}$ and mass $M_{\rm{cell}}$ at redshift $z$, the average number density of halos with mass $m$ that are virialized at redshift $z'$ can be calculated \citep{CoorySheth2002} from 
    \begin{eqnarray}
       n(m, z' \mid M_{\rm{cell}}, V_{\rm{cell}}, z) = f(\nu_{10})\frac{\Omega_m \rho_{\mathrm{crit}}}{m}\frac{d\nu_{10}}{dm}.
    \end{eqnarray}
    Here
    \begin{eqnarray}
        \nu_{10}=\frac{\left[\delta_{sc}(z_1)-\delta_0(\delta,z_0)\right]^2}{\sigma^2(m)-\sigma^2(M_{\mathrm{cell}})}, 
    \end{eqnarray}
    where $\rho_\mathrm{crit}$ represents the critical density, $\sigma(m)$ is the variance of density fluctuations on mass scale $m$ and $\delta_{\text{sc}}(z_1)$ denotes the critical density for spherical collapse at redshift $z_1$ extrapolated to the present using linear theory. The term $\delta_0(\delta, z_0)$ corresponds to the initial density necessary for a region to reach $\delta$ at redshift $z_0$. Following the approach of ref.~\cite{xu_yuebindm2hi1h_2019}, we utilize the form of $f(v)$ as described by ref.~\cite{sheth_large-scale_1999}. To obtain a realistic subgrid halo distribution, we introduce a Poisson fluctuation such that $n_p V_{\rm{cell}}  \sim \text{Poisson}(n V_{\rm{cell}}  )$. Ultimately, by applying the same halo mass-HI mass relation from ref.~\cite{Villaescusa-Navarro2018}, the HI mass in each cell is calculated by summing over all subgrid halo masses
    \begin{eqnarray}
    M_{\rm{HI,Subgrid}}(\text{cell}) = \sum_{m=M_{\rm{hard}}}^{M_{\rm{halo, res}}} n_p V_{\rm{cell}} M_{\rm{HI}}(m, z),
    \end{eqnarray}
    where $M_{\rm{hard}}$ is the hard cutoff mass from Table 1 of ref.~\cite{Villaescusa-Navarro2018}. 
    
From the HI distribution $M_{\rm HI}$, including both resolved halos and subgrid contributions, we then calculate the total 21cm temperature fluctuation as \citep{PrichardLoeb2012}:
\begin{equation}
    \delta T_{21} \approx 27 x_{\mathrm{HI}}(1+\delta_\mathrm{b}) \left(\frac{\Omega_b h^2}{0.023}\right) \left(\frac{0.15}{\Omega_m h^2} \frac{1+z}{10}\right)^{1/2},  
    \label{eq:prichard7}
\end{equation}
where $x_{\mathrm{HI}}$ is the neutral hydrogen fraction, and we have assumed the spin temperature is significantly higher than background radiation, $T_\mathrm{S} \gg T_\gamma$. The term $(1+\delta_\mathrm{b})$ is given by:
\begin{equation}
    1+\delta_\mathrm{b} = \frac{M_{\mathrm{HI}}/V_{\mathrm{cell}}}{f_\mathrm{H} 
 \Omega_{\mathrm{b}}\rho_\mathrm{crit}}  ,
\end{equation}
where $f_{\mathrm{H}} = 0.76$ is the hydrogen fraction. At low redshifts, HI is primarily concentrated in halos, so $x_{\mathrm{HI}}$ is set to unity.

\section{Impact of Instrumental Thermal Noise}\label{apdx:noise}

\rev{Instrumental noise is unavoidable in real observations, and here we evaluate its impact on the positivization process. For simplicity, the noise is generated as Gaussian random field in configuration space as a pixel-by-pixel independent realization, with its variance with its variance defined relative to the 21cm signal variance $\sigma_{\rm 21}$. The results show that phase restoration remains effective when the instrumental noise is comparable to or lower than the 21cm signal level. For instance, the correlation ratio at $k=k_{\rm cut}=0.5~h\,\mathrm{Mpc^{-1}}$ can decrease from approximately $0.9$ with no noise to around $0.7$ when $\sigma_{\rm noise} = \sigma_{\rm 21}$. However, performance deteriorates further as thermal noise significantly exceeds the signal level.}

\rev{
In addition to phase recovery, we also examine the impact on the power spectrum (panel b and d) and the two-point correlation function (panel c and e). As shown, increasing noise levels further suppress the positivized power spectra, likely due to the lack of consistent mode coupling of the noise contribution. This has significant implications, as it demands a larger rescaling factor to match a theoretical power spectrum, leading to even greater noise amplification. And a similar effect can be observed in panels (d) and (e), where the Fourier amplitude of the positivized field is rescaled using the same factor derived from the noiseless case.
Meanwhile, as shown in panel (c) and (e),  the measured correlation function becomes noticeably noisier, leading to the gradual washing out of BAO peaks. However, at a noise level comparable to the 21cm signal, the BAO peak remains identifiable, suggesting a degree of robustness in the positivization approach.  
}

\section{Numerical Optimization of Amplitudes}\label{apdx:amp_opt}

Since the positivization process only restores the phase angle of missing modes, we now explore the potential for further recovering the amplitudes through numerical optimization. For this purpose, we continue to utilize the non-negativity requirement. However, since the positivized field has already been `optimized' to eliminate all negative values in real space, additional operation are necessary to initiate the optimization process.  Knowing from previous analysis that the amplitude of positivized field is underestimated, we initially increase the amplitude slightly to begin the optimization and then iteratively minimize the occurrence of negative values in the configuration space. 

\begin{figure}
    \centering
    \includegraphics[width=1.0\linewidth]{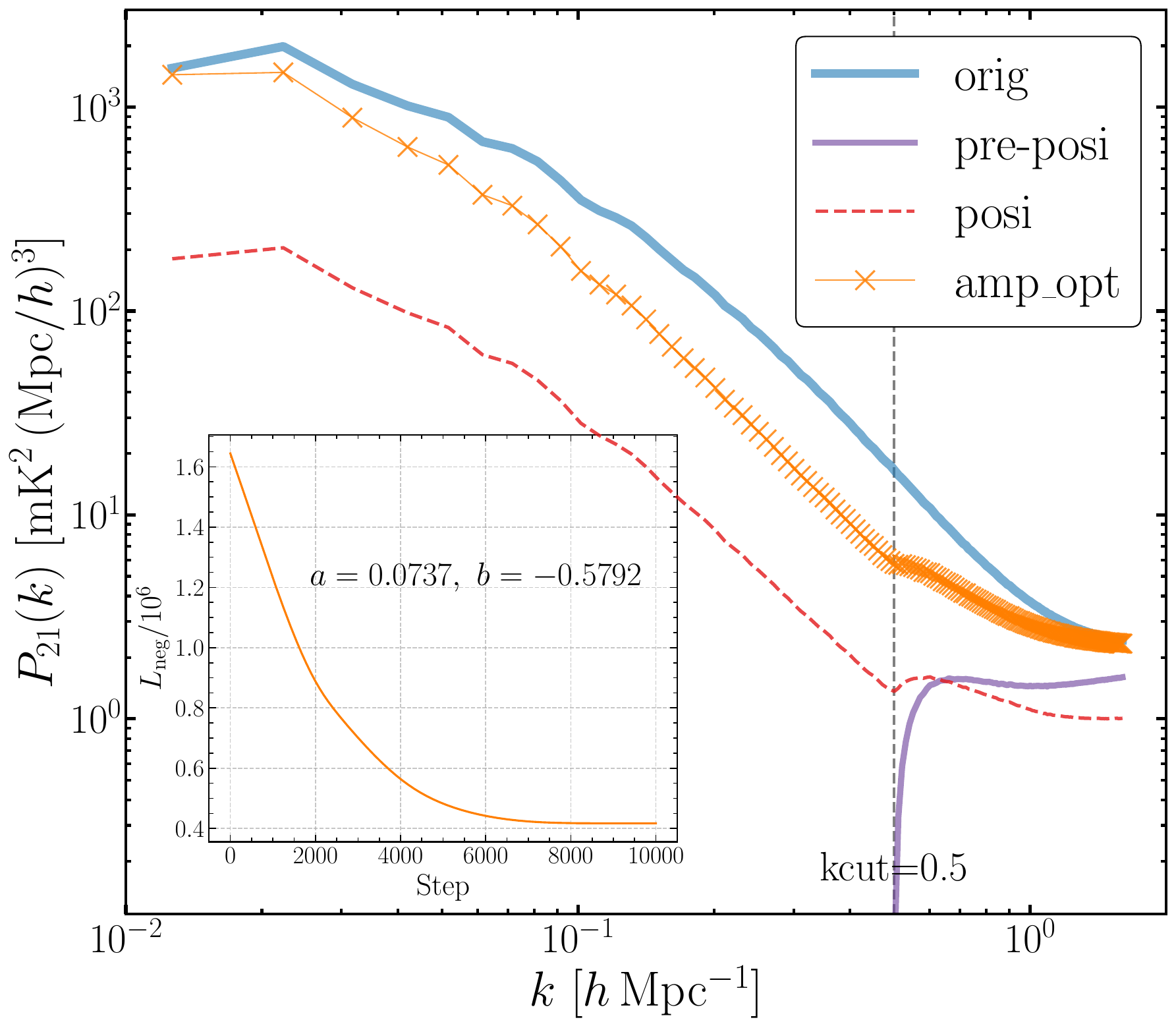}
    \caption{Comparison of the 21cm power spectra from various stages of data processing including the original field (blue solid line), the positivized result (red dashed line), the numerical optimized result (orange cross line), and the pre-positivized field (purple solid line). 
    The inset tracks the evolution of the loss function, Eq.~\ref{eq:loss}, across iteration steps.
    }
    \label{FIG:amp_opt}
\end{figure}

Specifically, we employ the gradient descent algorithm and define a loss function to penalize negative values in real space, given by
\begin{equation}
L_{\rm neg} =\sum_{\vx}\max\left (0, -T_{\rm 21}(\vx) \right). 
\label{eq:loss}
\end{equation}
Here the summation is performed over all real space location. We have also explored alternative definitions, such as counting only the number of negative values; however, Eq. \ref{eq:loss} has demonstrated the best performance in our tests. To effectively manage the extremely large parameter space of all missing amplitudes, we observe that the transfer function $\kappa(k) = P_{\mathrm{posi}}(k) / P_{\mathrm{orig}}(k)$, as shown in the panel (c) and (d) of  Fig. \ref{FIG:Ck1d}, is quite smooth. Consequently, as a first-order approximation, we model the log transfer function 
\begin{eqnarray}
    \log(\kappa)(k) = a \log(k) + b
    \label{eqn:tfmodel}
\end{eqnarray}
linearly in log-k space, thereby reducing the optimization to a two-dimensional parameter space $\boldsymbol{\alpha}=\{ a, b\}$. 
Additionally, we notice that among all degenerate solutions, the zeroth mode is challenging to constrain using only the non-negativity condition, as a sufficiently large $T_{\rm 21}(k=0)$ could always satisfy this condition. Therefore, in this study, we assume that the mean temperature at a specific redshift can be known, either through theoretical predictions or from other observational sources. Based on these considerations, our numerical algorithm incorporates the following key components: 
\begin{itemize}
    \item \textbf{Combined phases:} After positivization, while the phase angles of missing modes align with their true values, the process slightly alters the phases of remaining modes. Consequently, we merge the restored missing phases with the original phases of the remaining modes.     
    \item \textbf{Optimizing amplitude transfer function:} To expedite the optimization process, and considering the smooth relation between the positivized and true power spectrum, we choose to optimize the transfer function $\kappa(k)$ using a linear model, rather than adjusting all Fourier amplitudes individually.
    \item \textbf{Fixed zeroth mode:} Since the mean temperature $\langle \delta T_{21} \rangle$ can be set high enough to satisfy the non-negativity requirement, we assume it is known.
\end{itemize}

In practical implementation, we iteratively update parameters $\boldsymbol{\alpha}=\{ a, b \}$ with the following rule 
\begin{eqnarray}
    \boldsymbol{\alpha}_{n+1} =   \boldsymbol{\alpha}_{n} - \eta \partial_{\boldsymbol{\alpha}} L_{\rm neg}
\end{eqnarray}
to minimize the loss function $L_{\rm neg}$, where $\eta$ represents the learning rate. The gradient of each step is calculated using the automatic numerical differentiation engine, autograd, from \texttt{PyTorch} \citep{NEURIPS2019_bdbca288}. 
For optimization, we employ the Adaptive Moment Estimation (Adam) optimizer \citep{kingma2014adam}, which integrates the benefits of momentum and adaptive learning rates. This optimizer dynamically adjusts the learning rate for each parameter, improving both the efficiency and flexibility of the training process. The iterative updates continue until the algorithm converges or achieves a zero-loss condition.

The optimization result is displayed in Fig. \ref{FIG:amp_opt}. Here, we have set a global learning rate of $10^{-4}$, with the initial values for $a$ and $b$ both starting at $0$. Upon convergence, the optimal values were determined to be $a=0.0737$ and $b=-0.5792$, and the sum of the negative regions in real space is approximately $4.2 \times 10^5$ (as depicted in Fig.~\ref{FIG:amp_opt}). As illustrated in the figure, our simple two-parameter optimization approach has successfully increased the amplitude by an order of magnitude (orange cross line, labeled as `amp\_opt'),  compared to the positivized field (red dashed line). This enhancement brings the amplitude to within a factor of a few of the true amplitude (thick blue solid line). 
Additionally, we illustrate in panel (d) of Fig.~\ref{FIG:BAO} the correlation function of the amplitude-optimized field (blue cross line, labeled `amp\_opt'). Compared to the positivized field (shown in panel c and notice for its much smaller scale), the correlation function of the optimized field aligns more closely in scale with the true field. However, the BAO peak appears somewhat smeared, indicating that improved amplitude restoration is necessary for more accurate BAO measurements.


\bibliography{ms}{}

\begin{thebibliography}{}
\expandafter\ifx\csname natexlab\endcsname\relax\def\natexlab#1{#1}\fi
\providecommand{\url}[1]{\href{#1}{#1}}
\providecommand{\dodoi}[1]{doi:~\href{http://doi.org/#1}{\nolinkurl{#1}}}
\providecommand{\doeprint}[1]{\href{http://ascl.net/#1}{\nolinkurl{http://ascl.net/#1}}}
\providecommand{\doarXiv}[1]{\href{https://arxiv.org/abs/#1}{\nolinkurl{https://arxiv.org/abs/#1}}}

\bibitem[{Amiri(2022)}]{chime2022}
Amiri, M. 2022, The Astrophysical Journal Supplement Series

\bibitem[{{Amiri} {et~al.}(2023){Amiri}, {Bandura}, {Chen}, {Deng}, {Dobbs},
  {Fandino}, {Foreman}, {Halpern}, {Hill}, {Hinshaw}, {H{\"o}fer}, {Kania},
  {Landecker}, {MacEachern}, {Masui}, {Mena-Parra}, {Milutinovic},
  {Mirhosseini}, {Newburgh}, {Ordog}, {Pen}, {Pinsonneault-Marotte}, {Polzin},
  {Reda}, {Renard}, {Shaw}, {Siegel}, {Singh}, {Vanderlinde}, {Wang}, {Wiebe},
  {Wulf}, \& {CHIME Collaboration}}]{2023ApJ...947...16A}
{Amiri}, M., {Bandura}, K., {Chen}, T., {et~al.} 2023, \apj, 947, 16,
  \dodoi{10.3847/1538-4357/acb13f}

\bibitem[{{B{\'e}gin} {et~al.}(2022){B{\'e}gin}, {Liu}, \&
  {Gorce}}]{2022PhRvD.105h3503B}
{B{\'e}gin}, J.-M., {Liu}, A., \& {Gorce}, A. 2022, \prd, 105, 083503,
  \dodoi{10.1103/PhysRevD.105.083503}

\bibitem[{Behroozi {et~al.}(2013)Behroozi, Wechsler, \& Wu}]{rockstar2013}
Behroozi, P.~S., Wechsler, R.~H., \& Wu, H.-Y. 2013, The Astrophysical Journal,
  762, 109, \dodoi{10.1088/0004-637X/762/2/109}

\bibitem[{Bernardeau {et~al.}(2002)Bernardeau, Colombi, Gaztañaga, \&
  Scoccimarro}]{Bernardeau2002PTreview}
Bernardeau, F., Colombi, S., Gaztañaga, E., \& Scoccimarro, R. 2002, Physics
  Reports, 367, 1, \dodoi{https://doi.org/10.1016/S0370-1573(02)00135-7}

\bibitem[{{Bigot-Sazy} {et~al.}(2015){Bigot-Sazy}, {Dickinson}, {Battye},
  {Browne}, {Ma}, {Maffei}, {Noviello}, {Remazeilles}, \&
  {Wilkinson}}]{2015MNRAS.454.3240B}
{Bigot-Sazy}, M.~A., {Dickinson}, C., {Battye}, R.~A., {et~al.} 2015, \mnras,
  454, 3240, \dodoi{10.1093/mnras/stv2153}

\bibitem[{{Chang} {et~al.}(2010){Chang}, {Pen}, {Bandura}, \&
  {Peterson}}]{Chang2010Natur}
{Chang}, T.-C., {Pen}, U.-L., {Bandura}, K., \& {Peterson}, J.~B. 2010, Nature,
  466, 463, \dodoi{10.1038/nature09187}

\bibitem[{Chapman {et~al.}(2012)Chapman, Abdalla, Harker, Jelić, Labropoulos,
  Zaroubi, Brentjens, de~Bruyn, \& Koopmans}]{Chapman2012}
Chapman, E., Abdalla, F.~B., Harker, G., {et~al.} 2012, Monthly Notices of the
  Royal Astronomical Society, 423, 2518,
  \dodoi{10.1111/j.1365-2966.2012.21065.x}

\bibitem[{Cooray \& Sheth(2002)}]{CoorySheth2002}
Cooray, A., \& Sheth, R. 2002, Physics Reports, 372, 1,
  \dodoi{10.1016/S0370-1573(02)00276-4}

\bibitem[{{Ding} {et~al.}(2024){Ding}, {Wang}, {Pen}, \&
  {Li}}]{2024ApJS..274...44D}
{Ding}, J., {Wang}, X., {Pen}, U.-L., \& {Li}, X.-D. 2024, \apjs, 274, 44,
  \dodoi{10.3847/1538-4365/ad6f0a}

\bibitem[{{Drinkwater} {et~al.}(2010){Drinkwater}, {Jurek}, {Blake}, {Woods},
  {Pimbblet}, {Glazebrook}, {Sharp}, {Pracy}, {Brough}, {Colless}, {Couch},
  {Croom}, {Davis}, {Forbes}, {Forster}, {Gilbank}, {Gladders}, {Jelliffe},
  {Jones}, {Li}, {Madore}, {Martin}, {Poole}, {Small}, {Wisnioski}, {Wyder}, \&
  {Yee}}]{Drinkwater2009}
{Drinkwater}, M.~J., {Jurek}, R.~J., {Blake}, C., {et~al.} 2010, Monthly
  Notices of the Royal Astronomical Society, 401, 1429,
  \dodoi{10.1111/j.1365-2966.2009.15754.x}

\bibitem[{{Gagnon-Hartman} {et~al.}(2021){Gagnon-Hartman}, {Cui}, {Liu}, \&
  {Ravanbakhsh}}]{2021MNRAS.504.4716G}
{Gagnon-Hartman}, S., {Cui}, Y., {Liu}, A., \& {Ravanbakhsh}, S. 2021, \mnras,
  504, 4716, \dodoi{10.1093/mnras/stab1158}

\bibitem[{{Kennedy} {et~al.}(2024){Kennedy}, {Carr}, {Gagnon-Hartman}, {Liu},
  {Mirocha}, \& {Cui}}]{2024MNRAS.529.3684K}
{Kennedy}, J., {Carr}, J.~C., {Gagnon-Hartman}, S., {et~al.} 2024, \mnras, 529,
  3684, \dodoi{10.1093/mnras/stae760}

\bibitem[{Kingma \& Ba(2014)}]{kingma2014adam}
Kingma, D.~P., \& Ba, J. 2014, arXiv preprint arXiv:1412.6980,
  \dodoi{https://doi.org/10.48550/arXiv.1412.6980}

\bibitem[{{Lewis} \& {Challinor}(2011)}]{2011ascl.soft02026L}
{Lewis}, A., \& {Challinor}, A. 2011, {CAMB: Code for Anisotropies in the
  Microwave Background}, Astrophysics Source Code Library, record ascl:1102.026

\bibitem[{{Li} {et~al.}(2024){Li}, {Wang}, {Li}, {Ding}, {Luan}, \&
  {Luo}}]{LiQian2024arXiv}
{Li}, Q., {Wang}, X., {Li}, X., {et~al.} 2024, arXiv e-prints,
  arXiv:2412.04021, \dodoi{10.48550/arXiv.2412.04021}

\bibitem[{{Liu} \& {Shaw}(2020)}]{2020PASP..132f2001L}
{Liu}, A., \& {Shaw}, J.~R. 2020, \pasp, 132, 062001,
  \dodoi{10.1088/1538-3873/ab5bfd}

\bibitem[{{Liu} {et~al.}(2009){Liu}, {Tegmark}, {Bowman}, {Hewitt}, \&
  {Zaldarriaga}}]{Liua2009MNRAS}
{Liu}, A., {Tegmark}, M., {Bowman}, J., {Hewitt}, J., \& {Zaldarriaga}, M.
  2009, Monthly Notices of the Royal Astronomical Society, 398, 401,
  \dodoi{10.1111/j.1365-2966.2009.15156.x}

\bibitem[{{Makinen} {et~al.}(2021){Makinen}, {Lancaster},
  {Villaescusa-Navarro}, {Melchior}, {Ho}, {Perreault-Levasseur}, \&
  {Spergel}}]{2021JCAP...04..081M}
{Makinen}, T.~L., {Lancaster}, L., {Villaescusa-Navarro}, F., {et~al.} 2021,
  \jcap, 2021, 081, \dodoi{10.1088/1475-7516/2021/04/081}

\bibitem[{{Mangena} {et~al.}(2020){Mangena}, {Hassan}, \&
  {Santos}}]{2020MNRAS.494..600M}
{Mangena}, T., {Hassan}, S., \& {Santos}, M.~G. 2020, \mnras, 494, 600,
  \dodoi{10.1093/mnras/staa750}

\bibitem[{Maniyar {et~al.}(2021)Maniyar, {Ali-Ha{\"i}moud}, Carron, Lewis, \&
  Madhavacheril}]{Maniyar2021}
Maniyar, A.~S., {Ali-Ha{\"i}moud}, Y., Carron, J., Lewis, A., \& Madhavacheril,
  M.~S. 2021, PHYS. REV. D

\bibitem[{Masui {et~al.}(2013)Masui, Switzer, Banavar, Bandura, Blake, Calin,
  Chang, Chen, Li, Liao, Natarajan, Pen, Peterson, Shaw, \& Voytek}]{Masui2013}
Masui, K.~W., Switzer, E.~R., Banavar, N., {et~al.} 2013, The Astrophysical
  Journal, 763, L20, \dodoi{10.1088/2041-8205/763/1/L20}

\bibitem[{{Newburgh} {et~al.}(2016){Newburgh}, {Bandura}, {Bucher}, {Chang},
  {Chiang}, {Cliche}, {Dav{\'e}}, {Dobbs}, {Clarkson}, {Ganga}, {Gogo},
  {Gumba}, {Gupta}, {Hilton}, {Johnstone}, {Karastergiou}, {Kunz}, {Lokhorst},
  {Maartens}, {Macpherson}, {Mdlalose}, {Moodley}, {Ngwenya}, {Parra},
  {Peterson}, {Recnik}, {Saliwanchik}, {Santos}, {Sievers}, {Smirnov},
  {Stronkhorst}, {Taylor}, {Vanderlinde}, {Van Vuuren}, {Weltman}, \&
  {Witzemann}}]{HIRAX2016SPIE}
{Newburgh}, L.~B., {Bandura}, K., {Bucher}, M.~A., {et~al.} 2016, in Society of
  Photo-Optical Instrumentation Engineers (SPIE) Conference Series, Vol. 9906,
  Ground-based and Airborne Telescopes VI, ed. H.~J. {Hall}, R.~{Gilmozzi}, \&
  H.~K. {Marshall}, 99065X, \dodoi{10.1117/12.2234286}

\bibitem[{Paszke {et~al.}(2019)Paszke, Gross, Massa, Lerer, Bradbury, Chanan,
  Killeen, Lin, Gimelshein, Antiga, Desmaison, Kopf, Yang, DeVito, Raison,
  Tejani, Chilamkurthy, Steiner, Fang, Bai, \& Chintala}]{NEURIPS2019_bdbca288}
Paszke, A., Gross, S., Massa, F., {et~al.} 2019, in Advances in Neural
  Information Processing Systems, ed. H.~Wallach, H.~Larochelle,
  A.~Beygelzimer, F.~d~Alch\'{e}-Buc, E.~Fox, \& R.~Garnett, Vol.~32 (Curran
  Associates, Inc.).
\newblock
  \url{https://proceedings.neurips.cc/paper_files/paper/2019/file/bdbca288fee7f92f2bfa9f7012727740-Paper.pdf}

\bibitem[{Paul {et~al.}(2023)Paul, Santos, Chen, \& Wolz}]{Paul2023}
Paul, S., Santos, M.~G., Chen, Z., \& Wolz, L. 2023, A First Detection of
  Neutral Hydrogen Intensity Mapping on {{Mpc}} Scales at
  \$z{\textbackslash}approx 0.32\$ and \$z{\textbackslash}approx 0.44\$,
  arXiv.
\newblock \doarXiv{2301.11943}

\bibitem[{{Paul} {et~al.}(2023){Paul}, {Santos}, {Chen}, \&
  {Wolz}}]{Paul23Meerkat}
{Paul}, S., {Santos}, M.~G., {Chen}, Z., \& {Wolz}, L. 2023, arXiv e-prints,
  arXiv:2301.11943, \dodoi{10.48550/arXiv.2301.11943}

\bibitem[{Petrovic \& Oh(2011)}]{Petrovic2011}
Petrovic, N., \& Oh, S.~P. 2011, Monthly Notices of the Royal Astronomical
  Society, 413, 2103, \dodoi{10.1111/j.1365-2966.2011.18276.x}

\bibitem[{{Planck Collaboration} {et~al.}(2020){Planck Collaboration}, Aghanim,
  Akrami, Ashdown, Aumont, Baccigalupi, Ballardini, Banday, Barreiro, Bartolo,
  Basak, Battye, Benabed, Bernard, Bersanelli, Bielewicz, Bock, Bond, Borrill,
  Bouchet, Boulanger, Bucher, Burigana, Butler, Calabrese, Cardoso, Carron,
  Challinor, Chiang, Chluba, Colombo, Combet, Contreras, Crill, Cuttaia,
  De~Bernardis, De~Zotti, Delabrouille, Delouis, Di~Valentino, Diego, Dor{\'e},
  Douspis, Ducout, Dupac, Dusini, Efstathiou, Elsner, En{\ss}lin, Eriksen,
  Fantaye, Farhang, Fergusson, {Fernandez-Cobos}, Finelli, Forastieri, Frailis,
  Fraisse, Franceschi, Frolov, Galeotta, Galli, Ganga, {G{\'e}nova-Santos},
  Gerbino, Ghosh, {Gonz{\'a}lez-Nuevo}, G{\'o}rski, Gratton, Gruppuso,
  Gudmundsson, Hamann, Handley, Hansen, Herranz, Hildebrandt, Hivon, Huang,
  Jaffe, Jones, Karakci, Keih{\"a}nen, Keskitalo, Kiiveri, Kim, Kisner, Knox,
  Krachmalnicoff, Kunz, {Kurki-Suonio}, Lagache, Lamarre, Lasenby, Lattanzi,
  Lawrence, Le~Jeune, Lemos, Lesgourgues, Levrier, Lewis, Liguori, Lilje,
  Lilley, Lindholm, {L{\'o}pez-Caniego}, Lubin, Ma, {Mac{\'i}as-P{\'e}rez},
  Maggio, Maino, Mandolesi, Mangilli, {Marcos-Caballero}, Maris, Martin,
  Martinelli, {Mart{\'i}nez-Gonz{\'a}lez}, Matarrese, Mauri, McEwen, Meinhold,
  Melchiorri, Mennella, Migliaccio, Millea, Mitra, {Miville-Desch{\^e}nes},
  Molinari, Montier, Morgante, Moss, Natoli, {N{\o}rgaard-Nielsen}, Pagano,
  Paoletti, Partridge, Patanchon, Peiris, Perrotta, Pettorino, Piacentini,
  Polastri, Polenta, Puget, Rachen, Reinecke, Remazeilles, Renzi, Rocha,
  Rosset, Roudier, {Rubi{\~n}o-Mart{\'i}n}, {Ruiz-Granados}, Salvati, Sandri,
  Savelainen, Scott, Shellard, Sirignano, Sirri, Spencer, Sunyaev, {Suur-Uski},
  Tauber, Tavagnacco, Tenti, Toffolatti, Tomasi, Trombetti, Valenziano,
  Valiviita, Van~Tent, Vibert, Vielva, Villa, Vittorio, Wandelt, Wehus, White,
  White, Zacchei, \& Zonca}]{planck2020}
{Planck Collaboration}, Aghanim, N., Akrami, Y., {et~al.} 2020, Astronomy \&
  Astrophysics, 641, A6, \dodoi{10.1051/0004-6361/201833910}

\bibitem[{Pritchard \& Loeb(2012)}]{PrichardLoeb2012}
Pritchard, J.~R., \& Loeb, A. 2012, Reports on Progress in Physics, 75, 086901,
  \dodoi{10.1088/0034-4885/75/8/086901}

\bibitem[{{Sabti} {et~al.}(2024){Sabti}, {Reddy}, {Mu{\~n}oz}, {Mishra-Sharma},
  \& {Youn}}]{2024arXiv240721097S}
{Sabti}, N., {Reddy}, R., {Mu{\~n}oz}, J.~B., {Mishra-Sharma}, S., \& {Youn},
  T. 2024, arXiv e-prints, arXiv:2407.21097, \dodoi{10.48550/arXiv.2407.21097}

\bibitem[{{Santos} {et~al.}(2005){Santos}, {Cooray}, \& {Knox}}]{Santos2005ApJ}
{Santos}, M.~G., {Cooray}, A., \& {Knox}, L. 2005, The Astrophysical Journal,
  625, 575, \dodoi{10.1086/429857}

\bibitem[{{Shaw} {et~al.}(2014){Shaw}, {Sigurdson}, {Pen}, {Stebbins}, \&
  {Sitwell}}]{RShaw2014ApJ}
{Shaw}, J.~R., {Sigurdson}, K., {Pen}, U.-L., {Stebbins}, A., \& {Sitwell}, M.
  2014, \apj, 781, 57, \dodoi{10.1088/0004-637X/781/2/57}

\bibitem[{{Shaw} {et~al.}(2015){Shaw}, {Sigurdson}, {Sitwell}, {Stebbins}, \&
  {Pen}}]{RShaw2015PhRvD}
{Shaw}, J.~R., {Sigurdson}, K., {Sitwell}, M., {Stebbins}, A., \& {Pen}, U.-L.
  2015, \prd, 91, 083514, \dodoi{10.1103/PhysRevD.91.083514}

\bibitem[{Sheth \& Tormen(1999)}]{sheth_large-scale_1999}
Sheth, R.~K., \& Tormen, G. 1999, Monthly Notices of the Royal Astronomical
  Society, 308, 119, \dodoi{10.1046/j.1365-8711.1999.02692.x}

\bibitem[{Springel {et~al.}(2021)Springel, Pakmor, Zier, \&
  Reinecke}]{gadget2021}
Springel, V., Pakmor, R., Zier, O., \& Reinecke, M. 2021, Monthly Notices of
  the Royal Astronomical Society, 506, 2871, \dodoi{10.1093/mnras/stab1855}

\bibitem[{{Square Kilometre Array Cosmology Science Working Group}
  {et~al.}(2020){Square Kilometre Array Cosmology Science Working Group},
  {Bacon}, {Battye}, {Bull}, {Camera}, {Ferreira}, {Harrison}, {Parkinson},
  {Pourtsidou}, {Santos}, {Wolz}, {Abdalla}, {Akrami}, {Alonso},
  {Andrianomena}, {Ballardini}, {Bernal}, {Bertacca}, {Bengaly}, {Bonaldi},
  {Bonvin}, {Brown}, {Chapman}, {Chen}, {Chen}, {Cunnington}, {Davis},
  {Dickinson}, {Fonseca}, {Grainge}, {Harper}, {Jarvis}, {Maartens}, {Maddox},
  {Padmanabhan}, {Pritchard}, {Raccanelli}, {Rivi}, {Roychowdhury},
  {Sahl{\'e}n}, {Schwarz}, {Siewert}, {Viel}, {Villaescusa-Navarro}, {Xu},
  {Yamauchi}, \& {Zuntz}}]{SKA2020PASA}
{Square Kilometre Array Cosmology Science Working Group}, {Bacon}, D.~J.,
  {Battye}, R.~A., {et~al.} 2020, Publications of the Astronomical Society of
  Australia, 37, e007, \dodoi{10.1017/pasa.2019.51}

\bibitem[{Switzer {et~al.}(2015)Switzer, Chang, Masui, Pen, \&
  Voytek}]{switzer2015}
Switzer, E.~R., Chang, T.-C., Masui, K.~W., Pen, U.-L., \& Voytek, T.~C. 2015,
  The Astrophysical Journal, 815, 51

\bibitem[{Switzer {et~al.}(2013)Switzer, Masui, Bandura, Calin, Chang, Chen,
  Li, Liao, Natarajan, Pen, Peterson, Shaw, \& Voytek}]{switzer2013}
Switzer, E.~R., Masui, K.~W., Bandura, K., {et~al.} 2013, Monthly Notices of
  the Royal Astronomical Society: Letters, 434, L46,
  \dodoi{10.1093/mnrasl/slt074}

\bibitem[{{Vanderlinde} {et~al.}(2019){Vanderlinde}, {Liu}, {Gaensler}, {Bond},
  {Hinshaw}, {Ng}, {Chiang}, {Stairs}, {Brown}, {Sievers}, {Mena}, {Smith},
  {Bandura}, {Masui}, {Spekkens}, {Belostotski}, {Dobbs}, {Turok}, {Boyle},
  {Rupen}, {Landecker}, {Pen}, \& {Kaspi}}]{2019clrp.2020...28V}
{Vanderlinde}, K., {Liu}, A., {Gaensler}, B., {et~al.} 2019, in Canadian Long
  Range Plan for Astronomy and Astrophysics White Papers, Vol. 2020, 28,
  \dodoi{10.5281/zenodo.3765414}

\bibitem[{{Villaescusa-Navarro} {et~al.}(2018){Villaescusa-Navarro}, Genel,
  Castorina, Obuljen, Spergel, Hernquist, Nelson, Carucci, Pillepich,
  Marinacci, Diemer, Vogelsberger, Weinberger, \&
  Pakmor}]{Villaescusa-Navarro2018}
{Villaescusa-Navarro}, F., Genel, S., Castorina, E., {et~al.} 2018, The
  Astrophysical Journal, 866, 135, \dodoi{10.3847/1538-4357/aadba0}

\bibitem[{{Villanueva-Domingo} \&
  {Villaescusa-Navarro}(2021)}]{2021ApJ...907...44V}
{Villanueva-Domingo}, P., \& {Villaescusa-Navarro}, F. 2021, \apj, 907, 44,
  \dodoi{10.3847/1538-4357/abd245}

\bibitem[{{Wadekar} {et~al.}(2021){Wadekar}, {Villaescusa-Navarro}, {Ho}, \&
  {Perreault-Levasseur}}]{2021ApJ...916...42W}
{Wadekar}, D., {Villaescusa-Navarro}, F., {Ho}, S., \& {Perreault-Levasseur},
  L. 2021, \apj, 916, 42, \dodoi{10.3847/1538-4357/ac033a}

\bibitem[{Wolz {et~al.}(2014)Wolz, Abdalla, Blake, Shaw, Chapman, \&
  Rawlings}]{wolz2014}
Wolz, L., Abdalla, F.~B., Blake, C., {et~al.} 2014, Monthly Notices of the
  Royal Astronomical Society, 441, 3271, \dodoi{10.1093/mnras/stu792}

\bibitem[{{Wolz} {et~al.}(2017){Wolz}, {Blake}, {Abdalla}, {Anderson}, {Chang},
  {Li}, {Masui}, {Switzer}, {Pen}, {Voytek}, \& {Yadav}}]{2017MNRAS.464.4938W}
{Wolz}, L., {Blake}, C., {Abdalla}, F.~B., {et~al.} 2017, \mnras, 464, 4938,
  \dodoi{10.1093/mnras/stw2556}

\bibitem[{Xu {et~al.}(2019)Xu, Xu, Yue, Iliev, Trac, Gao, \&
  Chen}]{xu_yuebindm2hi1h_2019}
Xu, W., Xu, Y., Yue, B., {et~al.} 2019, Monthly Notices of the Royal
  Astronomical Society, 490, 5739, \dodoi{10.1093/mnras/stz2926}

\bibitem[{{Xu} {et~al.}(2015){Xu}, {Wang}, \& {Chen}}]{Tianlai2015ApJ}
{Xu}, Y., {Wang}, X., \& {Chen}, X. 2015, The Astrophysical Journal, 798, 40,
  \dodoi{10.1088/0004-637X/798/1/40}

\bibitem[{Zhu {et~al.}(2016)Zhu, Pen, Yu, Er, \& Chen}]{zhu2016}
Zhu, H.-M., Pen, U.-L., Yu, Y., Er, X., \& Chen, X. 2016, Physical Review D,
  93, 103504, \dodoi{10.1103/PhysRevD.93.103504}

\end{thebibliography}
\bibliographystyle{aasjournal}



\end{document}